\documentclass[review]{elsarticle}

\usepackage{lineno,hyperref}
\usepackage{graphicx}
\usepackage{float}
\usepackage{amsmath}
\usepackage{amssymb}
\usepackage{tipa}
\usepackage{enumerate}
\usepackage{booktabs}
\usepackage{array}
\usepackage{colortbl}

\modulolinenumbers[5]

\journal{Journal of \LaTeX\ Templates}

\newlength{\onesixth}
\newcolumntype{C}{>{\begin{minipage}{2\onesixth}\begin{center}}{c}<{\end{center}\end{minipage}}}
\newcolumntype{D}{>{\begin{minipage}{3\onesixth}\begin{center}}{c}<{\end{center}\end{minipage}}}
\newcolumntype{E}{@{}l@{}}

\begin{document}

\begin{frontmatter}

\title{ Fast asynchronous updating algorithms for $k$-shell indices}

%% Group authors per affiliation:
\author[inst1,inst2]{Yan-Li Lee}
\author[inst1,inst2]{Tao Zhou\corref{cor1}}
\cortext[cor1]{Corresponding author.}
\ead{zhutou@ustc.edu}
\address[inst1]{CompleX Lab, Web Sciences Center, University of Electronic Science and Technology of China, Chengdu 611731, P.R. China}
\address[inst2]{Big Data Research Center, University of Electronic Science and Technology of China, Chengdu 611731, P.R. China}

\begin{abstract}
 Identifying influential nodes in networks is a significant and challenging task. Among many centrality indices, the $k$-shell index performs very well in finding out influential spreaders. However, the traditional method for calculating the $k$-shell indices of nodes needs the global topological information, which limits its applications in large-scale dynamically growing networks. Recently, L\@\"{u} \emph{et al.} [Nature Communications 7 (2016) 10168] proposed a novel asynchronous algorithm to calculate the $k$-shell indices, which is suitable to deal with large-scale growing networks. In this paper, we propose two algorithms to select nodes and update their intermediate values towards the $k$-shell indices, which can help in accelerating the convergence of the calculation of $k$-shell indices. The former algorithm takes into account the degrees of nodes while the latter algorithm prefers to choose the node whose neighbors' values have been changed recently. We test these two methods on four real networks and three artificial networks. The results suggest that the two algorithms can respectively reduce the convergence time up to 75.4\% and 92.9\% in average, compared with the original asynchronous updating algorithm.
\end{abstract}

\begin{keyword}
\texttt Complex networks\sep Vital nodes identification\sep K-shell index\sep Fast algorithms
\end{keyword}

\end{frontmatter}

\section{Introduction}
The booming of network science \cite{newman2010networks,boccaletti2006complex,barabasi2013network} gives rise to a lot of novel ideas and methods to biology \cite{barabasi2004network,barabasi2011network,ideker2012differential,csermely2013structure}, economics \cite{jackson2008social,schweitzer2009economic,haldane2011systemic,hidalgo2009building}, social science \cite{kumar2006structure,scott2012social,castellano2009statistical}, data science \cite{lu2011link,lu2012recommender,zanin2016combining}, and so on. Recently, the focus of network science has been shifting from revealing the macroscopic statistical regularities (e.g., scale-free \cite{barabasi1999emergence}, assortative mixing \cite{newman2002assortative}, small-world \cite{watts1998collective} and clustering \cite{watts1998collective}) to discovering the mecroscopic structural organization (communities \cite{newman2004finding,fortunato2010community} and motifs \cite{alon2007network,Benson163}), and then to distinguishing the roles played by individual nodes and links. In particular, the discovery of scale-free property implies the significance of identifying the influential nodes \cite{lu2016vital,kempe2003maximizing,bakshy2011everyone}. For example, vital disease-related genes can help diagnose the known diseases and understand the features of unknown diseases \cite{csermely2013structure,barabasi2011network}, essential spreaders assist us to better control the outbreak of epidemics \cite{pastor2002immunization, cohen2003efficient,pastor2015epidemic}, influential customers allow us to conduct a successful advertisements marketing with low cost \cite{leskovec2007dynamics,huang2012exploring}.

To identify influential nodes, scientists \cite{hirsch2005index} have applied many centrality measures, such as degree, H-index \cite{korn2009lobby} (originally proposed by Hirsch \cite{hirsch2005index, pastor2016topological}), betweenness \cite{ freeman1977set}, $k$-shell index \cite{kitsak2010identification} (also called $k$-core index or coreness), and the like. As the most widely used measure, degree centrality counts the number of the nearest neighbors, therefore the importances of the nodes with the same degree are treated identically. Kitsak \emph{et al.} \cite{kitsak2010identification} argued that the location of a node is more important than the number of the node's  nearest neighbors in evaluating the node's influence on the spreading dynamics, namely the nodes in the core position are more influential than the nodes in the periphery. And many studies show that $k$-shell index and its variants perform better than degree in the identification of influential nodes \cite{kitsak2010identification, pei2014searching, castellano2012competing, liu2015core, qing62new, luo2016identifying}, at least in the range of moderate infectivity for some spreading processes \cite{klemm2012measure,liu2016locating}.

The traditional algorithm uses $k$-core decomposition \cite{dorogovtsev2006k} to calculate the $k$-shell index. In a simple network, all the isolated nodes are assigned $k$-shell value $k_S=0$ and then removed from the network. Next it removes all the nodes with degree $k$=1, which probably leads to some new nodes with degree $k\leqslant1$. These nodes are removed again until all the remaining nodes are of degree $k\textgreater 1$. The removed nodes along with the links among them form the $1$-shell and the nodes' $k$-shell indices are $k_S$=1. The process continues to remove all the nodes with degree $k\leqslant2$ iteratively and all the nodes and links removed in this round constitute the $2$-shell. By analogy, we repeat this operation and ultimately every node will be assigned a $k_S$ value. The $k$-core decomposition process has to restart when a few nodes or a few links are added to the network, so it faces a tough challenge when being applied in large-scale dynamically growing networks.

Recently, L\@\"{u} \emph{et al.} \cite{lu2016h} proposed an \emph{asynchronous updating} algorithm to calculate the $k$-shell indices. In each step, it randomly selects one node to update its intermediate value towards the $k$-shell index by an operator $\mathcal{H}$. After convergence, the values of all nodes in the steady state are their $k$-shell indices. $\mathcal{H}$ is an operator on a group of real numbers $\left(x_1,x_2,\cdots,x_n\right)$, returning an integer $y$, which is the largest integer such that there are at least $y$ elements in $\left(x_1,x_2,\cdots,x_n\right)$ whose values are no less than $y$.

We have noticed that the random selection of nodes for updating will lead to slow convergence to $k$-shell indices, and thus we propose two heuristic algorithms to optimize the node selection strategy in the asynchronous updating process. One algorithm considers the degrees of nodes, which performs well in highly heterogeneous networks (the more uneven the degree distribution, the stronger the heterogeneity), whereas the other algorithm prefers to select the nodes whose neighbors' values have been changed recently. In order to demonstrate the advantage of our algorithms, we compare our algorithms with the original asynchronous updating algorithm \cite{lu2016h}. For a more adequate comparison, we further propose the so-called \emph{sequential asynchronous updating} algorithm as another baseline algorithm, which selects node in a specific order that can be prescribed arbitrarily. The numerical results on four real networks and three artificial networks indicate that our algorithms can remarkably fasten the convergence.

\section{Methods}
 Denote $G(V,E)$ a simple network, where $V$ is the set of nodes and $E$ is the set of links. The degree and $k$-shell index of an arbitrary node $i \in V$ are denoted as $k_i$ and $c_i$ respectively. The numbers of nodes and links are labeled as $\left|V\right| = N$ and $\left|E\right| = M$, respectively.

\subsection{Original asynchronous updating algorithm}

We call the original asynchronous updating algorithm \cite{lu2016h} as random asynchronous updating (RAU) algorithm. At each time step, RAU algorithm randomly selects a node $i$ and updates its $g$ value as:
\begin{equation}
g_i=\mathcal{H}\left(g_{j_1},g_{j_2},\cdots,g_{j_{k_i}}\right),
\end{equation}
where $j_1,j_2,\cdots,j_{k_i}$ are the nearest neighbors of node $i$, $g_{j_1},g_{j_2},\cdots,g_{j_{k_i}}$ are their current $g$ values and for each node $i$, its initial value is set as $g_i=k_i$. It has been proved that after the value $g_i$ of an arbitrary node $i$ converges to its $k$-shell index $c_i$, it will not be changed even if it is selected again \cite{lu2016h}.

\subsection{Sequential asynchronous updating algorithm}

Another straightforward method to select nodes is to follow a certain order of nodes, which is determined randomly in advance. For example, given an arbitrary order $v_1, v_2, \cdots, v_N$, the nodes being selected are $v_1, v_2, \cdots, v_N, v_1, v_2, \cdots, \\ v_N, v_1, \cdots$, until all values converge to $k$-shell indices. We name it as sequential asynchronous updating (SAU) algorithm. We will show later that SAU algorithm is a simple but efficient way in the calculation of $k$-shell values.

\subsection{Degree-Biased Algorithm}

Statistical result indicates that there is a positive association between degree and $k$-shell index and the distributions of them are all heterogeneous \cite{pastor2016topological,dorogovtsev2006k,lu2016h,carmi2007model,zhang2008evolution}. We suspect that nodes with different degrees may play different roles in the convergence process. Therefore we propose the degree-biased (DB) algorithm to select nodes by the ascending order of degrees (we also test the strategy of descending order, but it performs worse), which can be considered as a special case of the SAU algorithm.

\begin{figure}[ht]
    \centering
    \includegraphics[width=14cm]{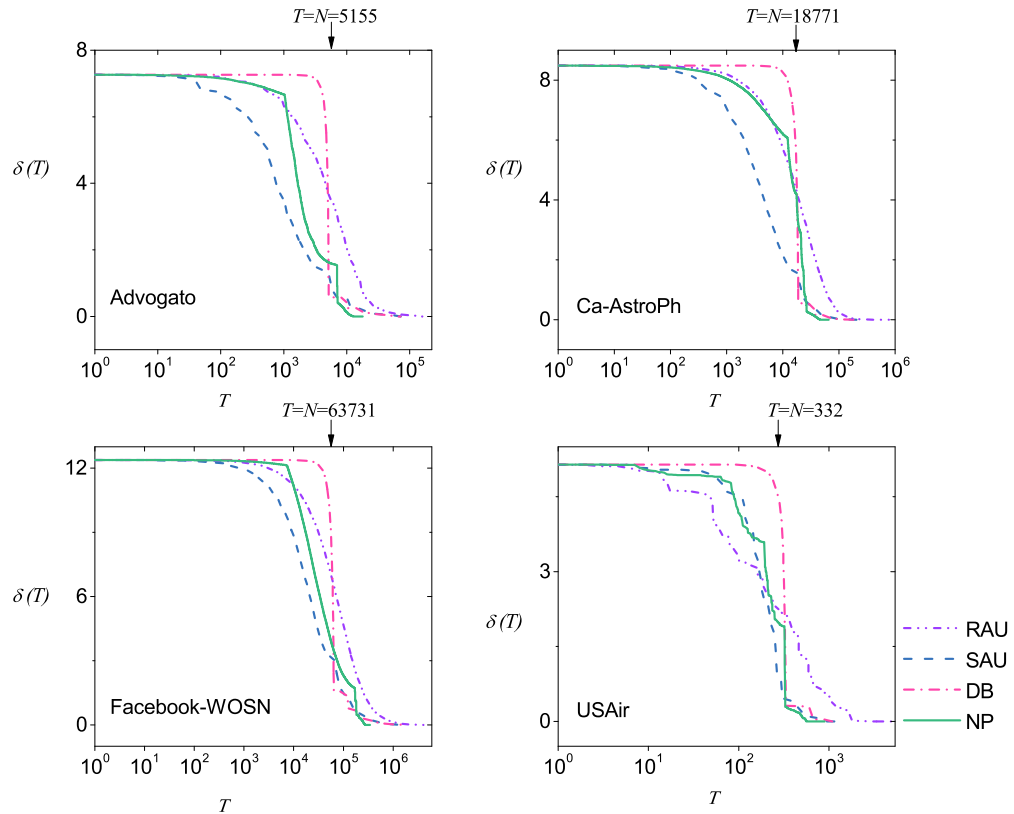}
    \centering
    \caption{$\delta(T)$ versus $T$ for the four algorithms on real networks.}
    \label{fig:fig1}
\end{figure}

\begin{figure}[ht]
    \centering
    \includegraphics[width=14cm]{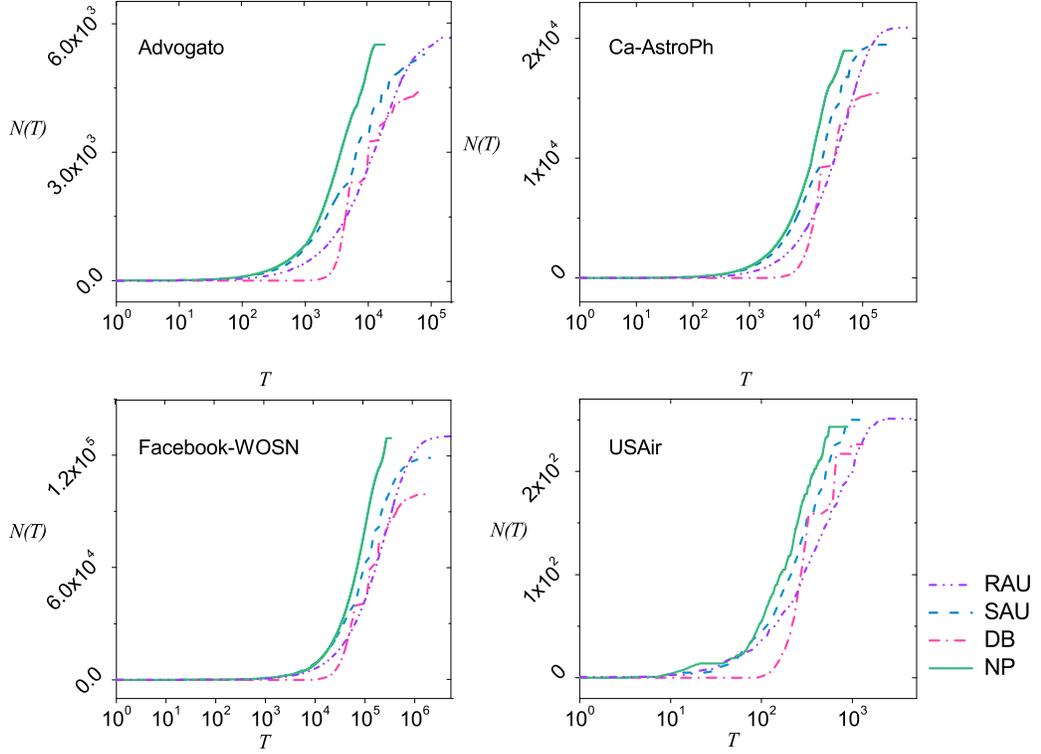}
    \centering
    \caption{$N(T)$ versus $T$ for the four algorithms on real networks.}
    \label{fig:fig2}
\end{figure}

\subsection{Neighborhood Preferential Algorithm}

Since the change of $g$ value of a node may induce further changes of its neighbors' $g$ values and so forth, we propose a neighborhood preferential (NP) algorithm. Similar to the SAU algorithm, we set up a random order of nodes, denoted by a circular sequence as $Q_s=\{v_1, v_2, \cdots, v_N, v_1, v_2, \cdots, v_N, v_1, \cdots \}$. Initially, a pointer is associated with the first node in $Q_s$, say $v_1$, and a set of nodes $Q_c$ is set as $Q_c=\emptyset$. The NP algorithm runs according to the following steps.

\emph{Step 1}. If $Q_c=\emptyset$, we select the node being pointed in $Q_s$ and update its $g$ value, and then move the pointer to the next node in $Q_s$. If $Q_c\neq \emptyset$, we randomly select a node from $Q_c$ and update its $g$ value, and then remove this node from $Q_c$.

\emph{Step 2}. Denote $v$, $g$ and $g'$ the node being selected in step 1, the $g$ value of $v$ before updating and the $g$ value of $v$ after updating. As being proved in \cite{lu2016h}, $g'\leq g$. If $g'=g$, return to step 1. If $g'<g$, go to step 3.

\emph{Step 3}. For each of $v$'s neighbor, say $v_j\in \Gamma_v$ where $\Gamma_v$ is the set of $v's$ neighbors, if $g\geq g_j$, $g'<g_j$ and $v_j$ is not in $Q_c$, then $v_j$ is added to $Q_c$. Go to step 1.

Notice that, the change of $v$'s $g$ value will further induce a change of its neighbor $v_j$'s $g$ value, only if the $g$ value of $v$ may contribute to $g_j$ before updating (i.e., $g\geq g_j$) and the $g$ value of $v$ cannot contribute to $g_j$ after updating (i.e., $g'<g_j$). This leads to the above condition in step 3. The algorithm terminates when every node's $g$ value converges.

\subsection{Evaluation Criteria}

The most intuitive way to quantify the performance of an algorithm is to count the average number of selections of a node until the convergence, which is denoted by $\left\langle n_\infty\right\rangle$. If the total number of selections is $n_\infty$, then
\begin{equation}
\left\langle n_\infty\right\rangle = \frac{1}{N}n_\infty.
\end{equation}

To understand the convergence process of algorithms, we quantify the distance from the current $g$ values to the converged state as:
\begin{equation}
\delta(T) = \frac{1}{N}\sum\limits_{i=1}^N(g_i(T)-c_i).
\end{equation}
When $\delta$=0, the corresponding algorithm converges.

Another considered quantity is the number of selections that lead to the nodes' $g$ values changed until step $T$, named as the effective selections, which is
\begin{equation}
N(T)=\sum\limits_{i=1}^N\sum\limits_{t=1}^Ta_i(t),
\end{equation}
where $a_i(t)=1$ if $g_i$ is changed at time step $t$, and $a_i(t)=0$ otherwise.

\begin{figure}[ht]
    \centering
    \includegraphics[trim=0mm 0mm 0mm 0mm,clip=true,width=14cm]{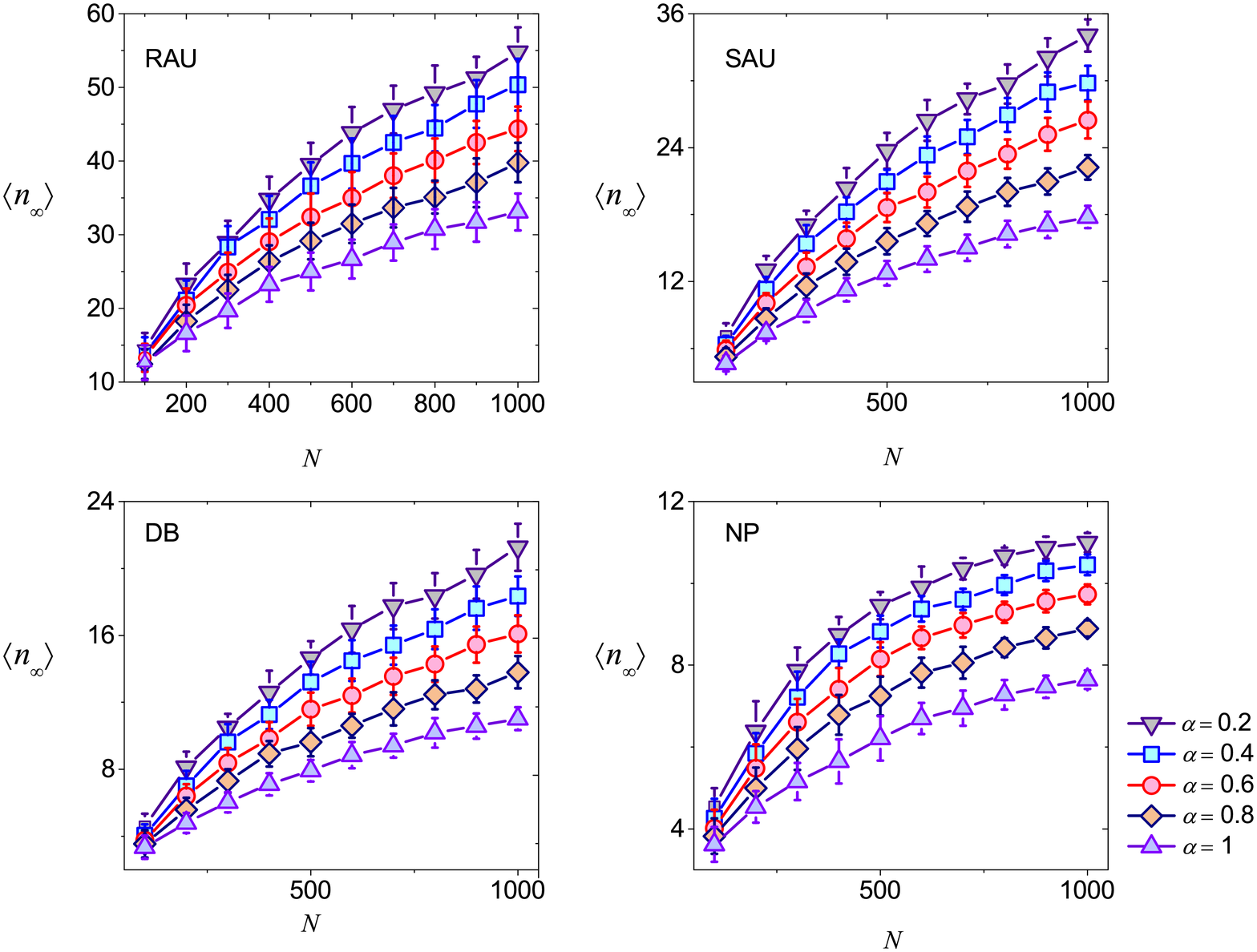}
    \centering
    \caption{The dependence of $\left\langle n_\infty \right\rangle$ on $\alpha$ and $N$ for the NPA model. Each data point is averaged over 50 independent runs and error bars represent standard deviations.}
    \label{fig:fig3}
\end{figure}

\section{Simulation Results}

\subsection{Real Networks}

\begin{table}[htbp]
\centering
\begin{tabular}{p{3.5cm}>{\centering}p{1cm}>{\centering}p{1cm}>{\centering}p{1cm}>{\centering}p{1cm}>{\centering}p{1cm}>{\centering\arraybackslash}p{0.8cm}}
\toprule
    Networks  &  $N$  &  $M$  &  $\left\langle k \right\rangle$  &  $\left\langle d \right\rangle$  &  $\left\langle c \right\rangle$  &  $r$\\
\midrule
\end{tabular}
\begin{tabular}{p{3.5cm}p{1cm}p{1cm}>{\raggedleft}p{0.5cm}@{.}p{0.3cm}>{\raggedleft}p{0.5cm}@{.}p{0.3cm}>{\raggedleft}p{0.5cm}@{.}p{0.3cm}>{\raggedleft}p{0.5cm}@{.}p{0.3cm}}
    \textbf{Real Networks} \\
    Advogato  & 5155 & 39285 & 15&24 & 3&27 & 0&319 & $\textnormal{-}$0&095 \\
    Ca-AstroPh & 18771 & 198050 & 21&10 & 4&19 & 0&677&0&205\\
    Facebook-WOSN & 63731 & 817035 & 25&64 & 4&32 & 0&253&0&177\\
    USAir & 332 & 2126 & 12&81 & 2&74 & 0&749 &$\textnormal{-}$0&208\\

    \textbf{Artificial Networks} \\
    ER & 1000 & 199935 & 399&87 & 1&60 & 0&400 &0&002\\
    NPA & 1000 & 9749 & 19&50 & 2&55 & 0&073&$\textnormal{-}$0&018\\
    WS & 1000 & 40000 & 80&00 & 1&93 & 0&151&0&004\\
\bottomrule
\end{tabular}
\caption{The basic topological features of the seven networks. $N$, $M$, $\left\langle k \right\rangle$, $\left\langle d \right\rangle$, $\left\langle c \right\rangle$ and $r$ are the number of nodes, the number of links, the average degree, the average distance, the average clustering coefficient \cite{watts1998collective} and the assortative coefficient \cite{newman2002assortative} respectively. The connecting probability of the ER model is set as $p$=0.4, the preferential attachment coefficient of the NPA model is set as $\alpha$=1, equivalent to the BA model, and the rewiring probability and the number of neighbors of the WS model are set as $q$=0.5 and $z$=40, respectively. The three artificial network examples are corresponding to the examples shown in Table 2.}
\label{Table:table1}
\end{table}

To compare the four asynchronous updating algorithms mentioned in this paper, we conduct experiments on four real networks drawn from disparate domains involving two social networks (Advogato and Facebook-WOSN), a coauthorship network (Ca-AstroPh) and a transportation network (USAir) (see Table \ref{Table:table1} for their basic statistics). Advogato (http://konect.uni-koblenz.de/networks
 /advogato) is a software declaration platform, where each developer represents a node and the trust relation from one developer to another developer represents a link. The data of Ca-AstroPh (http://snap.stanford.edu/data) is collected from January 1993 to April 2003 from the arXiv's Astrophysics section that covers 18771 authors and 198050 co-author relations. Facebook-WOSN (http://socialnetworks.mpi-sws.org/data-wosn2009.html) \cite{viswanath-2009-activity} was crawled through the Facebook New Orleans networks during the period from December 2008 to January 2009. A link is generated when a user's friend comment the user's wall. USAir (http://vlado.fmf.uni-lj.si/pub/networks/data/mix/USAir97.net) is a network of the US air transportation system that contains $332$ airports and $2126$ airlines. We remove all the multiple links, self-loops and the isolated nodes of the four networks. Meanwhile, the directed links are treated as undirected.

\subsection{Artificial Networks}
Three kinds of artificial networks are employed for comparison, including Erd{\"o}s-R{\'e}nyi (ER) networks \cite{erdos1959random,erd6s1960evolution,erdHos1961strength}, nonlinear preferential attachment (NPA) networks \cite{krapivsky2000connectivity,krapivsky2001organization,zhou2007modelling}, and Watts-Strogatz (WS) networks \cite{watts1998collective}. In an ER network, each pair of nodes is connected with a constant probability $p$. An NPA network starts from a fully connected network with $m_0$ nodes ($m_0$ is much smaller than the network size and thus the specific value of $m_0$ will not affect the statistical properties of NPA networks). At each time step we add a new node with $m$ links ($m\leqslant m_0$) to the existing nodes and the probability to connect to an existing node $v_i$ is proportional to $k_i^\alpha$. When the preferential exponent $\alpha$=1, the NPA model reduces to the well-known Barab\'asi-Albert (BA) model \cite{barabasi1999emergence}. WS network is initiated by a ring network with each node being connected to its $z$ nearest neighbors, then we rewire one end of each link to a randomly chosen node with probability $q$. In all the above models, multiple links and self-loops are not allowed. The basic statistics of the three artificial network examples are presented in Table \ref{Table:table1}, and the algorithmic performance for these three networks are shown in Table \ref{Table:table2}.

\subsection{Results}

\begin{table}[htbp]
\centering
\begin{tabular}{p{3.5cm}>{\centering\arraybackslash}p{1.2cm}>{\centering\arraybackslash}p{1.2cm}>{\centering\arraybackslash}p{1.2cm}>{\centering\arraybackslash}p{1.2cm}}
\toprule
    Networks  & RAU & SAU & DB & NP \\
\midrule
\end{tabular}

\begin{tabular}{p{3.5cm}>{\raggedleft}p{0.5cm}@{.}p{0.55cm}>{\raggedleft}p{0.5cm}@{.}p{0.55cm}>{\raggedleft}p{0.5cm}@{.}p{0.55cm}>{\raggedleft}p{0.5cm}@{.}p{0.55cm}}
 \textbf{Real Networks} \\
    Advogato  & 43&43 & 15&10 & 13&99 & \textbf{3}&\textbf{52}  \\
    Ca-AstroPh & 44&15 & 15&23 & 11&97 & \textbf{3}&\textbf{52} \\
    Facebook-WOSN  & 74&94 & 32&42 & 31&00 & \textbf{5}&\textbf{33} \\
    USAir & 15&96 & 3&76 & 3&93 & \textbf{2}&\textbf{67} \\

    \textbf{Artificial Networks} \\
    ER & 21&08 & 4&76 & \textbf{3}&\textbf{06} & 4&97\\
    NPA & 37&79 & 20&02 & 12&96 & \textbf{8}&\textbf{05}\\
    WS & 24&75 & 5&19 & 5&00 & \textbf{3}&\textbf{72}\\
\bottomrule
\end{tabular}
\caption{Algorithmic performance $\left\langle n_\infty \right\rangle$ for real and artificial networks, where the three artificial network examples are the same to those in Table \ref{Table:table1}. The best-performed results are emphasized in bold. }
\label{Table:table2}
\end{table}

The performances of the four algorithms, measured by $\left\langle n_\infty\right\rangle$, are summarized in Table \ref{Table:table2}. RAU performs worst while NP is overall the best. DB is the secondary best, which outperforms NP only for the example ER network.

Figure \ref{fig:fig1} exhibits the convergence processes of the four algorithms. For $\delta(T)$ of DB, a sharp drop appears in each plot when $T$ gets close to $N$, indicating that after a round of updates of low-degree nodes, the selections of large-degree nodes usually lead to remarkable changes of $g$ values, resulting in such observations. Figure \ref{fig:fig2} shows the number of effective selections versus time, where the total number of effective selections of DB in each plot is smallest. Given a network, the total change of $g$ values is the same under different algorithms, and thus the smallest $N(n_\infty)$ of DB suggests that if an updating causes a change of $g$ value, for DB, the change is bigger in average than the other three algorithms. This is also resulted from the quick drops of $g$ values of large-degree nodes, in accordance with the observation in Figure \ref{fig:fig1}. However, NP still performs better than DB since its mechanism guarantees that NP can produce more effective selections especially in the early stage.

Figure \ref{fig:fig3} reports the dependence of performance of the four algorithms on the parameters of the NPA model ($N$ and $\alpha$). Clearly, $\left\langle n_\infty\right\rangle$ is sensitive to the change of network topology under all the four algorithms, while NP performs the best under all situations. Explicitly speaking, as shown in Fig. 4, $\left\langle n_\infty\right\rangle$ increases monotonically as the increase of $N$ or the decrease of $\alpha$ for all considered algorithms. We also find the similar monotonic relationship between $\left\langle n_\infty\right\rangle$ and $N$ for ER and WS networks, but the relationships $\left\langle n_\infty\right\rangle \sim p$ and  $\left\langle n_\infty\right\rangle \sim q$ are irregular.

\begin{figure}[ht]
    \centering
    \includegraphics[trim=0mm 0mm 0mm 5mm,clip=true,width=14cm]{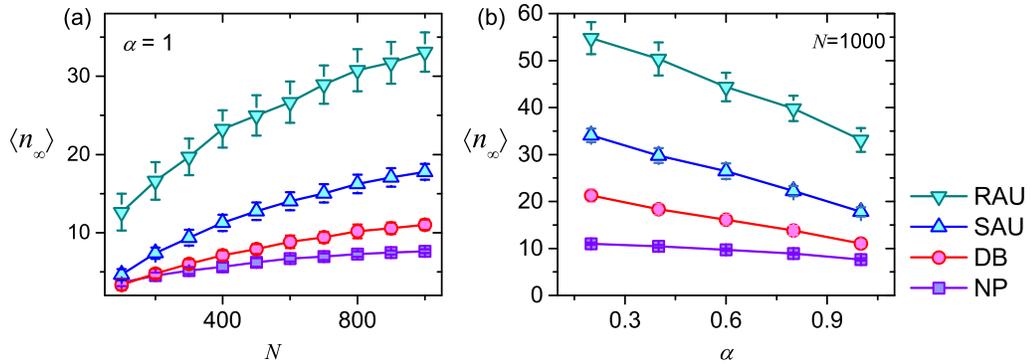}
    \centering
    \caption{ (a) $\left\langle n_\infty \right\rangle$ vs $N$. The plot exhibits a positive correlation between $\left\langle n_\infty \right\rangle$ and $N$ for the networks with fixed $\alpha$. (b) $\left\langle n_\infty \right\rangle$ vs $\alpha$. The plot exhibits a negative correlation between $\left\langle n_\infty \right\rangle$ and $\alpha$ for the networks with fixed $N$. Each data point is averaged over 50 independent runs and error bars represent standard deviations.}
    \label{fig:fig4}
\end{figure}

\section{Conclusion and Discussion}
In this paper, we proposed two novel algorithms, degree-biased algorithm and neighborhood preferential algorithm, to calculate $k$-shell indices of networks in an asynchronous way. Compared with the original algorithm \cite{lu2016h}, our algorithms are much more effective in terms of the convergence time. In particular, the neighborhood preferential algorithm can reduce the convergence time up to 92.9\% in average (see Table \ref{Table:table2}). As a consequence, we can largely facilitate the applications of $k$-shell index in large-scale dynamically growing networks that require distributed and asynchronous algorithms.

The success of the neighborhood preferential algorithm lies on the localization of the operator $\mathcal{H}$. As the current $g$ value of a node only depends on the $g$ values of its neighbors according to $\mathcal{H}$, a change of a node's $g$ value may induce cascading changes of its neighbors and so forth, leading to the high efficiency of the neighborhood preferential algorithm to produce effective selections, as indicated in Fig. \ref{fig:fig2}. Due to the increasing power of information technology, real networks, such as Internet, www and online social networks, become larger and larger, and thus to handle or even know the global topological structure of a network becomes harder and harder. Accordingly, distributed operators in network science and network engineering will be more significant in the near future while the centralized methods will be less feasible. Therefore, we would like to emphasize that the idea embedded in the neighborhood preferential algorithm may find many applications where some localized operators are iteratively used to dig out valuable information in a network. Lastly, the proposed methods can be easily extended to deal with directed and weighted networks.

\section*{Acknowledgments}

The authors acknowledge Dr. Qiang Dong, Dr. Qianming Zhang and Mr. Chao Fan for valuable discussion. This work is partially supported by National Natural Science Foundation of China under Grants Nos. 61433014.

\biboptions{numbers,sort&compress}

\end{document}